\documentclass[%
aps,
prl,
amsmath,amssymb,
preprint,
superscriptaddress
]{revtex4-1}

\usepackage{graphicx}
\usepackage{dcolumn}
\usepackage{bm}

\begin{document}

\title[Diffractive imaging of a transient core-shell nanoplasma]{Diffractive imaging of transient electronic core-shell structures in a nanoplasma}%

\author{Daniela Rupp}
\email[Corresponding author: ]{daniela.rupp@mbi-berlin.de}
\altaffiliation[Current address: ]{Max-Born-Institut Berlin, Max-Born-Stra{\ss}e 2A, 12489 Berlin, Germany}
\affiliation{IOAP, Technische Universit{\"a}t Berlin, Hardenbergstra{\ss}e 36, 10623 Berlin, Germany}
\author{Leonie Fl{\"u}ckiger}
\affiliation{IOAP, Technische Universit{\"a}t Berlin, Hardenbergstra{\ss}e 36, 10623 Berlin, Germany}
\affiliation{ARC Centre of Excellence for Advanced Molecular Imaging, La Trobe University, Bundoora, Victoria 3086, Australia}
\author{Marcus Adolph}%
\affiliation{IOAP, Technische Universit{\"a}t Berlin, Hardenbergstra{\ss}e 36, 10623 Berlin, Germany}
\author{Tais Gorkhover}%
\affiliation{IOAP, Technische Universit{\"a}t Berlin, Hardenbergstra{\ss}e 36, 10623 Berlin, Germany}
\affiliation{SLAC National Accelerator Laboratory, 2575 Sand Hill Road, Menlo Park, CA 94025, USA}
\author{Maria Krikunova}
\altaffiliation[Current address: ]{ELI Beamlines, Institute of Physics, Czech Academy of Science, Na Slovance 2, 182 21 Prague, Czech Republic}
\affiliation{IOAP, Technische Universit{\"a}t Berlin, Hardenbergstra{\ss}e 36, 10623 Berlin, Germany}
%\affiliation{ELI Beamlines, Institute of Physics, Czech Academy of Science, Na Slovance 2, 182 21 Prague, Czech Republic}
\author{Jan Philippe M{\"u}ller}
\affiliation{IOAP, Technische Universit{\"a}t Berlin, Hardenbergstra{\ss}e 36, 10623 Berlin, Germany}
\author{Maria M{\"u}ller}
\affiliation{IOAP, Technische Universit{\"a}t Berlin, Hardenbergstra{\ss}e 36, 10623 Berlin, Germany}
\author{Tim Oelze}
\affiliation{IOAP, Technische Universit{\"a}t Berlin, Hardenbergstra{\ss}e 36, 10623 Berlin, Germany}
\author{Yevheniy Ovcharenko}
\affiliation{IOAP, Technische Universit{\"a}t Berlin, Hardenbergstra{\ss}e 36, 10623 Berlin, Germany}
\affiliation{European XFEL, Holzkoppel 4, 22869 Schenefeld, Germany}
\author{Mario Sauppe}%
\affiliation{IOAP, Technische Universit{\"a}t Berlin, Hardenbergstra{\ss}e 36, 10623 Berlin, Germany}
\author{Sebastian Schorb}
\affiliation{IOAP, Technische Universit{\"a}t Berlin, Hardenbergstra{\ss}e 36, 10623 Berlin, Germany}
\author{David Wolter}%
\affiliation{IOAP, Technische Universit{\"a}t Berlin, Hardenbergstra{\ss}e 36, 10623 Berlin, Germany}
\author{Marion Harmand}%
\affiliation{FLASH, DESY, Notkestra{\ss}e 85, 22603 Hamburg, Germany}
\author{Rolf Treusch}%
\affiliation{FLASH, DESY, Notkestra{\ss}e 85, 22603 Hamburg, Germany}
\author{Christoph Bostedt}
\email[Corresponding author: ]{cbostedt@anl.gov}
\affiliation{Chemical Sciences and Engineering Division, Argonne National Laboratory, 9700 S. Cass Avenue, Argonne, IL 60439, USA}
\affiliation{Department of Physics and Astronomy, Northwestern University, 2145 Sheridan Road, Evanston, IL 60208, USA}
\author{Thomas M{\"o}ller}%
\affiliation{IOAP, Technische Universit{\"a}t Berlin, Hardenbergstra{\ss}e 36, 10623 Berlin, Germany}

\date{\today}

\begin{abstract}
We have recorded the coherent diffraction images of individual xenon clusters with intense extreme ultraviolet pulses to elucidate the influence of light-induced electronic changes on the diffraction pattern. Using the FLASH free-electron laser we tuned the wavelength to specific xenon atomic and ionic resonances. The data show the emergence of a transient core-shell structure within the otherwise homogeneous sample. Simulations indicate that ionization and nanoplasma formation result in a cluster shell with strongly altered refraction. The presented resonant scattering approach enables imaging of ultrafast electron dynamics on their natural time scale.
\end{abstract}
\pacs{36.40.-c}
\keywords{X-ray scattering, single particle imaging, rare-gas clusters}%

\maketitle

Intense femtosecond short-wavelength pulses from free-electron lasers (FELs) open new avenues to investigate transient states and ultrafast processes with unprecedented spatial and temporal resolution\,\cite{Feldhaus2013, Bostedt2016, Bencivenga2015, Yabashi2013}. Examples include diverse topics ranging from the first demonstration of femtosecond coherent diffractive imaging (CDI)\,\cite{Chapman2006} and the 3D characterization of isolated nanoparticles\,\cite{Barke2015} to the visualization of quantum vortices in helium droplets\,\cite{Gomez2014}, and non-equilibrium dynamics in molecules \cite{Minitti2015} and clusters\,\cite{Ferguson2016, Gorkhover2016, Flueckiger2016}.

Typical CDI efforts concentrate on retrieving the atomic structure or density distribution of the sample. Ultrafast photon induced changes to the sample electronic structure are mostly discussed in terms of ``damage" in both, experimental\,\cite{Nass2015} and theoretical\,\cite{Santra2011} approaches. However, the availability of intense short-wavelength pulses also yields tremendous opportunity to directly image electronic structure changes with high spatial resolution in a time-resolved manner. During the X-ray scattering process the photons interact with the electrons that are either tightly bound to the nuclei or delocalized in the valence states. In particular near absorption resonances, the X-ray scattering cross-sections depend sensitively on the energy of the incoming photon and the electronic structure of the sample\,\cite{Bostedt2012,vonKorffSchmising2014,Wu2016}.

In this letter we demonstrate how resonant elastic scattering can be used to directly image the spatial distribution of transient charge states in an evolving nanoplasma. As samples we use submicron-sized clusters that are simultaneously transformed to a highly excited nanoplasma and imaged with a single intense femtosecond FEL pulse. On the timescale of the pulse the position of the clusters is frozen in space and ionic motion in the generated nanoplasma can be neglected\,\cite{Rupp2016}. Nevertheless, we do observe modulations in the scattering patterns that scale with the FEL intensity and that are characteristic for core-shell structures. As they are independent from the geometric arrangement of the atoms in the cluster, we attribute these modulations to light-induced electronic structure changes which is supported by Mie calculations and Monte-Carlo simulations. The models indicate that the electronic core-shell structures exhibit surprisingly sharp boundaries that act akin to a transient mirror within the nanoplasma. The experiments show the potential of resonant coherent diffractive imaging for taking snapshots of ultrafast ionization dynamics or charge migration in complex samples with femtosecond time and nanometer spatial resolution.

The experiments were performed at the soft X-ray free-electron laser FLASH\,\cite{Ackermann2007} with extreme ultraviolet (XUV) pulses at 91\,eV photon energy. This energy matches the giant 4d resonance of neutral xenon\,\cite{GMayer1941} and some Xe charge states\,\cite{Ederer1964, Itoh2001, Andersen2009, Emmons2005, Aguilar2006, Bizau2000} as discussed in detail below. The FEL beam with $10^{13}$ photons per 100 femtosecond pulse was focused into a 20\,$\mu$m (FWHM) spot, reaching power densities up to $5\times10^{14}$\,W/cm$^2$. The pulses intersected a highly diluted jet of very large xenon clusters\,\cite{Rupp2014}. An adjustable piezo skimmer slit ensured that only one single cluster is present in the focus volume per FEL shot\,\cite{Rupp2012}. The scattering patterns were measured with a previously described\,\cite{Bostedt2010, Bostedt2012} large area scattering detector. The size of each single, mostly spherical cluster could be determined from the spacing of the extrema in the diffraction patterns\,\cite{Rupp2014}. Within the size regime of $R=(400\pm50)$\,nm a total of 94 diffraction images were obtained. In addition to the diffraction images, coincident single-shot ion spectra were recorded\,\cite{Gorkhover2012,Rupp2016}.

\begin{figure}[t]
\includegraphics[width=0.48\textwidth,keepaspectratio=true]{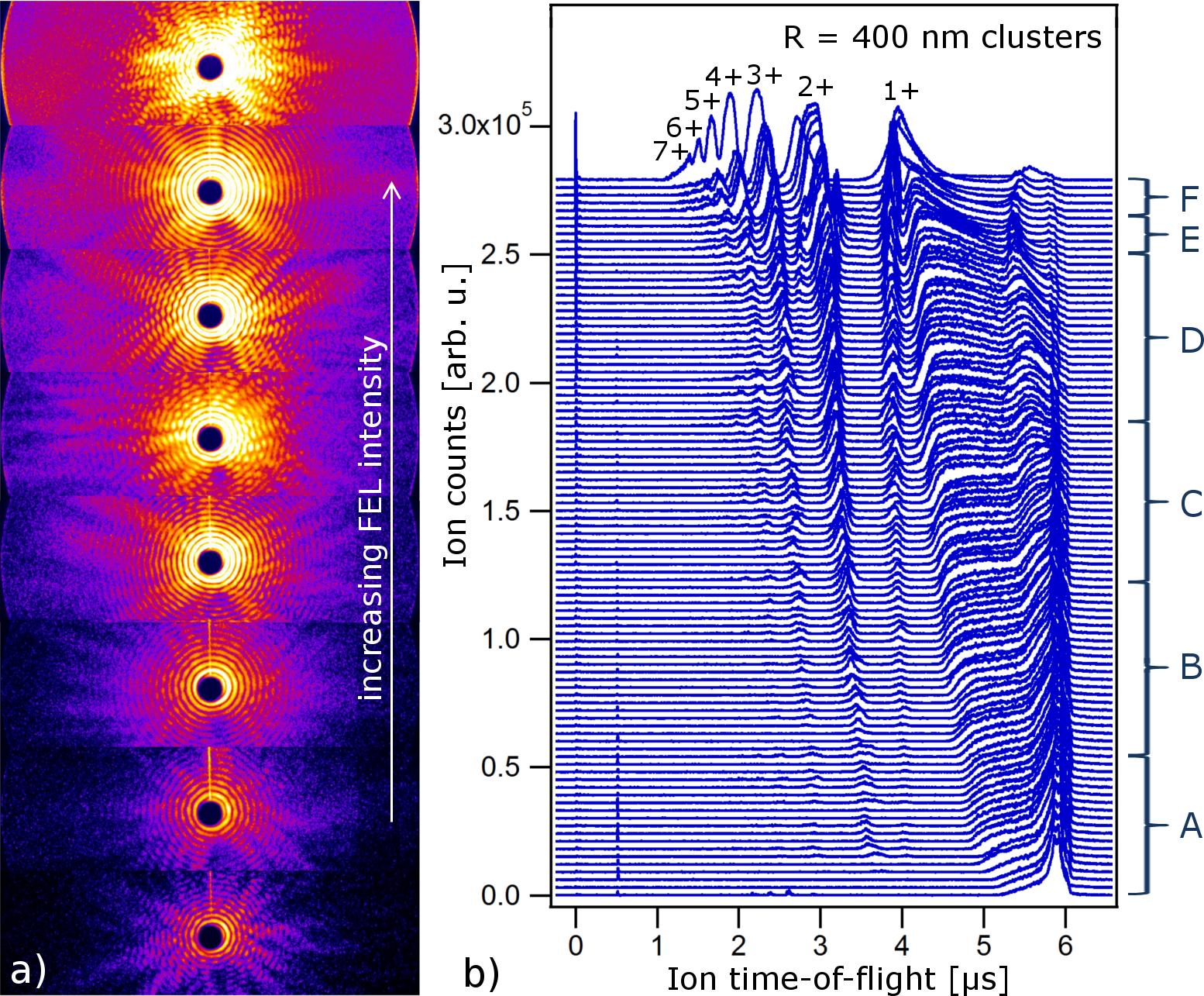}
\caption{\label{AlleBilder}Isolated xenon clusters were irradiated with intense XUV pulses (91\,eV photon energy, $5\times 10^{14}$\,W/cm$^2$ peak intensity). All events with single clusters of $(400\pm50)$\,nm radius were selected for analysis (total of 94 events) by the characteristic spacing of the diffraction rings. a) Representative diffraction images, intercepted at different positions of the focal power density distribution. b) Corresponding ion spectra of all 94 events (shifted by an offset for better visibility). The average kinetic energy of the $Xe^{1+}$ ions was used to sort the events for FEL exposure power density\,\cite{Rupp2016}. For the following analysis, all events were binned into categories A to F according to the abundance of higher charge states.}
\end{figure}

In Fig.\,\ref{AlleBilder}a examples for the single-shot diffraction images are displayed. The data were sorted for increasing FEL exposure intensity by the kinetic energy of the $Xe^{1+}$ ions from the time-of-flight spectra\,\cite{Rupp2016}, as shown in Fig.\,\ref{AlleBilder}b. In order to analyze only the intensity dependent changes in the patterns and to cancel out effects from irregular shapes and slightly different sizes, the data were binned into six categories A to F related to the appearance of the next higher charge states (see brackets on the right side of Fig.\,\ref{AlleBilder}b) and the diffraction patterns were averaged.

\begin{figure*}
\includegraphics[width=0.95\textwidth,keepaspectratio=true]{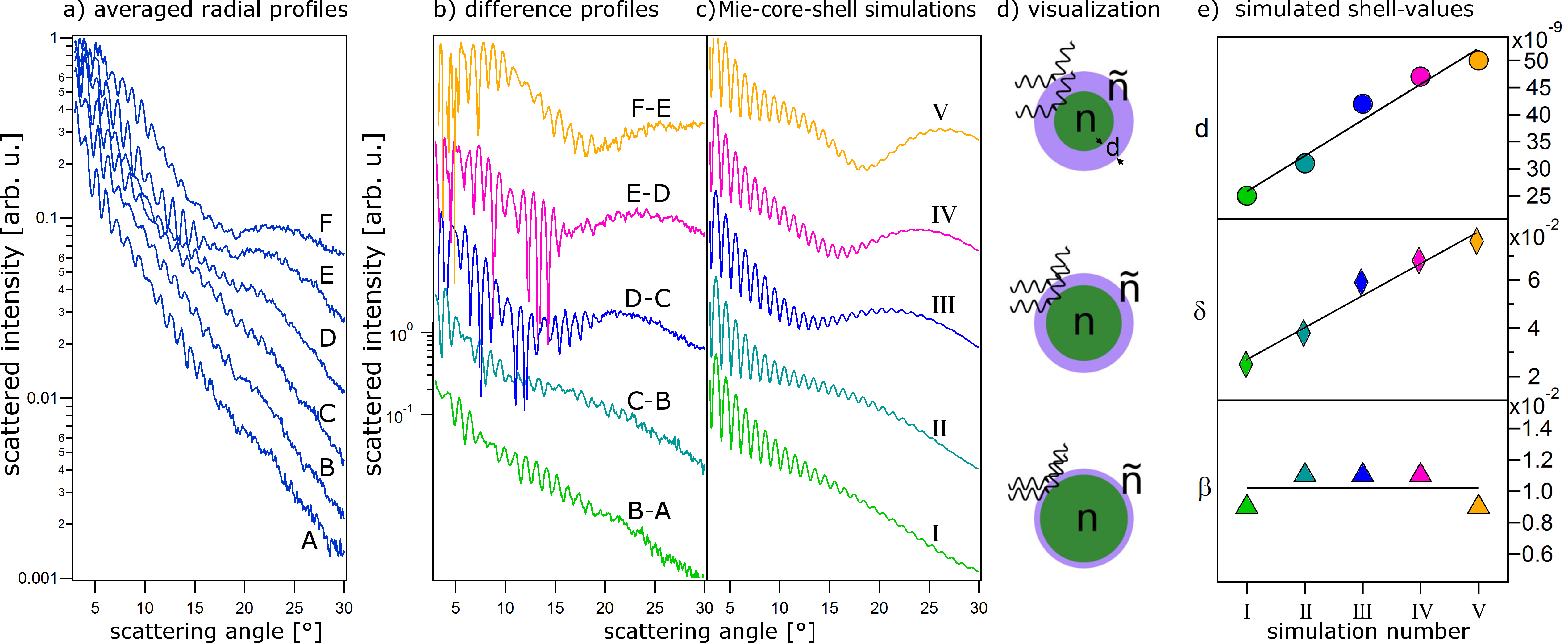}
\caption{\label{MieFits}a) Radial profiles of averaged diffraction images from bins A to F, as indicated in Fig.\,\ref{AlleBilder}b. The measured scattering intensities are corrected for the flat detector geometry\,\cite{Bostedt2012} and a nonlinear detector response\,\cite{Barke2015, Rupp2017}, radially averaged and plotted logarithmically vs. scattering angle. Profile A exhibits a ripple structure corresponding to the cluster size on an otherwise linearly dropping (log scale) curve, as expected from Mie theory for a homogeneous xenon cluster. Towards higher exposure power densities up to profile F, an additional large-scale structure evolves. b) Difference profiles from a). For better visibility the curves were shifted by multiplication with a factor. c) Matching calculated profiles obtained with a Mie-based core-shell code\,\cite{ShenCode}. d) Schematic visualization of the spatially inhomogeneous nanoplasma evolving with increasing FEL exposure power density. e) Parameters of the shell used as input for the calculations displayed in c), i.e. shell thickness $d$ (in m), refractive index decrement $\delta$, and absorption index $\beta$ (dimensionless, with $n=1-\delta+$\textbf{i}$\beta$). Black lines serve as guide to the eye. The refractive index of the core was kept constant to $n=1.007+$\textbf{i}$\cdot0.044$ (neutral xenon at 91\,eV\,\cite{HenkeTables}).}
\end{figure*}

The radial profiles resulting from thus averaged scattering images of each category A to F are displayed in Fig.\,\ref{MieFits}a. The high-frequency modulation of the profiles reflects the cluster size information. The envelope of profile A, representing the class of clusters exposed to lowest intensities, agrees well with the expected curve for a homogeneous spherical xenon cluster, dropping linearly on a logarithmic scale. In the absence of light induced changes in the particle, the profiles from clusters irradiated with higher FEL intensity would follow a similar curve, just with a proportionally higher scattering signal. In contrast, the profile envelopes B to F develop a more and more pronounced lobe structure roughly at $15^{\circ}$ to $30^{\circ}$ scattering angle. This evolving superstructure corresponds to the development of an additional characteristic length scale in the sample. In a classical Mie model, the lobe feature is characteristic for a core-shell structure with strongly deviating refractive indices in the shell compared to the core\,\cite{Mie1908,Bohren1983}. We apply a core-shell model to the data to extract estimates for the thickness and optical constants of the shell. Subsequently, we develop a physical picture of the plasma formation and discuss a possible origin of such a core-shell system as well as the limitations of this model.

In our Mie calculations we made some simplifying assumptions, namely that the profiles A to F, obtained at different FEL-intensities, reflect the course of the same evolution, but up to different stages. Individual steps of the evolution can therefore be extracted from the difference between each two profiles. This approach is conceptually similar to resonant imaging of magnetic domains and their ultrafast switching, where diffraction patterns are subtracted above and below an absorption edge\,\cite{vonKorffSchmising2014}. In Fig.\,\ref{MieFits}b the difference profiles from Fig.\,\ref{MieFits}a are given (F-E, E-D, and so on). The difference spectra reveal very distinct features in the superstructure: With increasing FEL intensity a broad lobe appears that becomes more and more pronounced, narrows, and shifts towards higher scattering angles.

Simulated profiles matching the superstructure in Fig.\,\ref{MieFits}b are shown in Fig.\,\ref{MieFits}c. The simulations were carried out using a code based on Mie theory and extended for spheres with a core-shell structure\,\cite{ShenCode, Mie1908, Bohren1983}. A schematic visualization in Fig.\,\ref{MieFits}d illustrates this evolution. The parameters of the shell used in these simulations are given in Fig.\,\ref{MieFits}e, specifically increasing shell thickness, increasing refractive index decrement $\delta$ (the refractive index $n$ is given by $n=1-\delta+$\textbf{i}$\beta$), and a very low absorption coefficient $\beta$ (factor 4 less than neutral xenon). Please note that while the tendencies found via the Mie simulations are probably correct, the absolute values might not match the actual optical constants in the nanoplasma because (i) the refractive index of the core is unknown, (ii) the nanoplasma evolution is continuous whereas we consider only five snapshots, and (iii) the nanoplasma structure may considerably deviate from a concentric core-shell system. In fact, it can be rather expected to be asymmetric in the direction of incident light\,\cite{Peltz2014}. Nevertheless, the good agreement between Fig.\,\ref{MieFits}b and c supports the hypothesis of a strongly altered outer shell in the cluster nanoplasma. It is notable that the core-shell structure appears to be a general feature because it survives the averaging over many single-cluster patters which themselves incorporate the average scattering signal over the FEL pulse duration. This raises the question of the origin of this refractive core shell system, i.e. the generation of a tens of nanometers thick shell of the nanoplasma with optical properties that differ so drastically from the plasma core.

\begin{figure}[b]
    \includegraphics[width=0.44\textwidth,keepaspectratio=true]{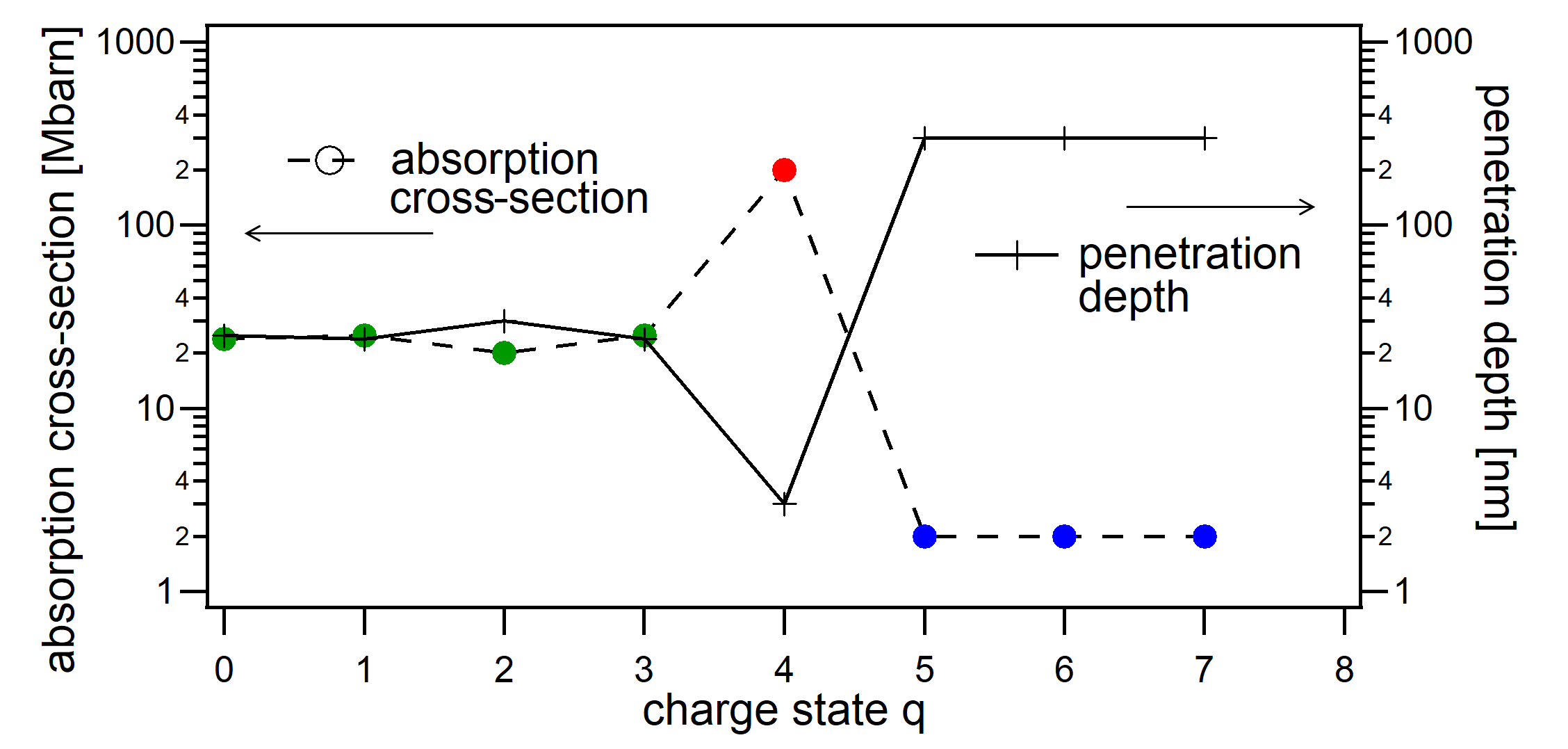}
    \caption{Absorption of xenon at 91\,eV. Total absorption cross-sections in Mbarn of neutral Xe\,\cite{Ederer1964} and Xe charge states 1+\,\cite{Itoh2001}, 2+\,\cite{Andersen2009}, 3+\,\cite{Emmons2005}, 4+\,\cite{Aguilar2006}, 5,6,7+\,\cite{Bizau2000} (colored points). Related penetration depth in nm (black crosses).}
    \label{XeProperties}
\end{figure}

\begin{figure}[t]
\includegraphics[width=0.35\textwidth,keepaspectratio=true]{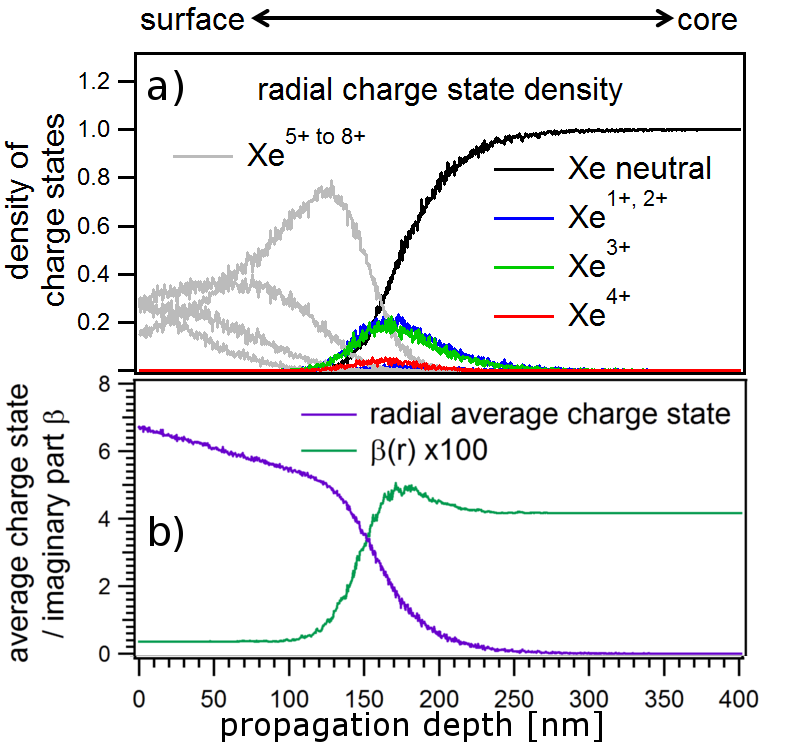}%Figure4b}
\caption{\label{ChargeStateDensity} a) Simulation of the radially changing charge state densities for a 400\,nm radius cluster irradiated with $1\cdot10^{14}W/cm^2$, corresponding to approximately 870 photons falling on the geometric cross-section of one xenon atom and propagating into the cluster from surface to core. Absorption cross-sections from Fig.\,3 are used to calculate the radial charge state densities. Only linear photo absorption is taken into account, while nonlinear effects, light scattering and plasma processes such as collisional ionization are neglected. b) The average charge state as derived from the radial charge state distributions is given in purple as a function of propagation depth $r$. Via $\beta=\frac{1}{4\pi}\lambda n_a \sigma_{abs}$\,\cite{Attwood} also the imaginary part of the refractive index $\beta(r)$ can be calculated (wavelength $\lambda$, $n_a$ atomic density\,\cite{Sears1962}, $\sigma_{abs}$ absorption cross-section).}
\end{figure}

The presumed origin of the core-shell structure lies in the peculiar electronic properties of xenon atoms and ions in the vicinity of the photon energy of 91\,eV. Absorption cross-sections for xenon atoms and atomic ions have been measured\,\cite{Ederer1964, Itoh2001, Andersen2009, Emmons2005, Aguilar2006, Bizau2000} and are summarized in Fig.\,\ref{XeProperties}. There is a clear step from high to low absorption between $Xe^{4+}$ and $Xe^{5+}$ with extremely high values for the charge state 4+ which exhibits a large ionic resonance at 91\,eV. Correspondingly, the penetration depth increases from about 30 to 300\,nm. To further investigate this observation, we model a first-order picture of the radial charge state distributions for our experiment using a Monte Carlo approach. The propagation of photons into the clusters is calculated starting at the cluster surface. At every atom or ion the absorption of the photon is tested, using probabilities according to the atomic and ionic cross-sections. The resulting radial charge state density distributions for an FEL intensity of $1\times10^{14}$\,W/cm$^2$ are given in Fig.\,\ref{ChargeStateDensity}a. The simple simulation allows to derive the average charge state as a function of the propagation depth and even to calculate the imaginary part of the refractive index $\beta$ (cf. caption of Fig.\,\ref{ChargeStateDensity}). Both curves given in Fig.\,\ref{ChargeStateDensity}b show that an outer shell exists up to a propagation depth of about 120\,nm that is exclusively populated by high charge states. Then the average charge state drastically drops in a transition region of about 100\,nm while the inner part of the cluster remains neutral. The absorption coefficient $\beta$ reveals an even more pronounced kink within only 50\,nm, resulting in an optical core-shell system with a rather transparent shell and opaque core.

Our simple atomistic model is in good agreement with the results from the Mie-calculations, showing the same trend however not the same absolute shell thicknesses (cf. Fig.\,\ref{MieFits}). Further, it is noted that the step in $\beta$ in Fig.\,\ref{ChargeStateDensity}b is still to soft to explain the pronounced superstructure in the scattering intensity. To generate pronounced modulations, two preconditions are necessary: (i) a transparent outer shell (low absorption index) and (ii) \emph{a sharp boundary} between regions of different refractive index. We have tested that the modulations would vanish in case of a rather smooth transition over several tens of nanometers as in Fig.\,4b. The required sharp transition is puzzling. Moreover, our atomistic model is expected to underestimate the transition regime width, as it neglects impact ionization and other charge transfer dynamics within the cluster that would smear out the charge state distribution.\\

However, the model only describes the absorption of the nanoplasma, i.e. the imaginary part $\beta$, and its radial dependence. For the optical response of the cluster both, the real and imaginary part of the refractive index $\delta$ and $\beta$ are relevant, which are interrelated through the Kramers-Kronig dispersions relations. In particular, a peak in $\beta$ (absorption resonance) is accompanied by a zero transition of $\delta$. Plasma calculations of the atomic scattering factors of $Xe^{3+}$\,\cite{Nilsen2007} indeed indicate that between 90\,eV and 98\,eV the real part of the atomic scattering factor $f_1$ (proportional to the refractive index decrement $\delta$) rapidly changes from strongly positive to negative values and back several times. Based on this information, the following consideration may provide an explanation for the required sharp change in refraction. We have to expect, that the atomic/ionic resonances are shifted by the plasma in the cluster\,\cite{HauRiegeDegenerate}, possibly up to several eV\,\cite{Gets2006}. Now considering the radial position of the $Xe^{3+}$-distribution in Fig.\,\ref{ChargeStateDensity}a (green curve) and comparing it with the average charge states at the same radial positions (purple curve) clearly shows that the environment of $Xe^{3+}$ ions strongly changes as a function of the propagation depth. Such a change in the plasma environment may translate into a radially dependent plasma shift of the electronic resonances up to several eV\,\cite{HauRiegeDegenerate}, from just below a sharp resonance to just above the resonance. This would result in a drastic change of the real part of the refractive index within a short distance, acting like a transient plasma mirror. A similar argument could be made for the xenon charge states 4+ to 6+, which also exhibit narrow and very strong absorption resonances in the vicinity of 91\,eV \cite{Emmons2005, Aguilar2006}. To test this hypothesis, a full description of the light propagation via sophisticated theoretical approaches will be needed that include impact ionization, charge transfer, plasma shifts of the energy levels, and further nanoplasma dynamics \cite{Saalmann2010, Peltz2012,Ho2017}. Nevertheless, these general considerations provide a first step towards understanding the observed results.

In summary, we have presented scattering patterns of single large xenon clusters resonantly excited with intense XUV pulses. The patterns reveal strong intensity dependent modulations of the scattering distribution, characteristic for core-shell systems. Mie-based simulations support a model of light induced electronic core-shell structures in the initially homogeneous systems with an increasingly thick outer shell characterized by low absorption and a strong and rapid change in refraction. The origin of this abrupt change in refractive index can be correlated to the radially changing plasma environment of higher charge states, translating into a radially changing shift of the electronic resonances. Our work shows that ultrafast light scattering can map the transient spatial charge distributions of resonant electronic states on the nanoscale. This method can be employed to develop a deeper understanding of nanoplasma formation and charge transfer dynamics which play a key role in many areas ranging from single-shot X-ray imaging to fusion and warm dense matter research as well as condensed matter physics. In the future, the approach provides an avenue to resolve ultrafast electron dynamics in extended systems on their natural time scale with intense attosecond pulses currently under development at FELs and lab-based sources \cite{Zholents2005,Saldin2006,Takahashi2013,Rupp2017}.

The authors thank Thomas Fennel and Christian Peltz for enlightening discussions. Excellent support from IOAP and DESY machine shops is acknowledged. The experiments have received funding from BMBF (grants 05K10KT2/05K13KT2) and DFG (grants MO 719/13-1 and /14-1). Further support is acknowledged from the U.S. Department of Energy, Office of Science, Basic Energy Sciences, Chemical Sciences, Geosciences, and Biosciences Division (C.B.), the Australian Research Council Centre of Excellence in Advanced Molecular Imaging, CE140100011 (L.F.), and the Volkswagen Foundation via a Peter Ewald fellowship (T.G.).

\providecommand{\noopsort}[1]{}\providecommand{\singleletter}[1]{#1}%

\end{document}